\documentclass[conference]{IEEEtran}
\IEEEoverridecommandlockouts
\usepackage{cite}
\usepackage{amsmath,amssymb,amsfonts}
\usepackage{algorithmic}
\usepackage{algorithm}
\usepackage{graphicx}
\usepackage{subfigure}
\usepackage{graphicx}
\usepackage{textcomp}
\usepackage{xcolor}
\def\BibTeX{{\rm B\kern-.05em{\sc i\kern-.025em b}\kern-.08em
    T\kern-.1667em\lower.7ex\hbox{E}\kern-.125emX}}
\begin{document}

\title{Energy Disaggregation \&
Appliance Identification in a Smart Home: Transfer Learning enables Edge Computing \\
}

\author{
\IEEEauthorblockN{
M. Hashim Shahab\IEEEauthorrefmark{1}, Hasan Mujtaba Buttar\IEEEauthorrefmark{1}, Ahsan Mehmood\IEEEauthorrefmark{1}, Waqas Aman\IEEEauthorrefmark{2}, M. Mahboob Ur Rahman\IEEEauthorrefmark{1},\\ M. Wasim Nawaz\IEEEauthorrefmark{3}, Haris Pervaiz\IEEEauthorrefmark{4},  Qammer H. Abbasi\IEEEauthorrefmark{5},\IEEEauthorrefmark{6} }

\IEEEauthorblockA{\IEEEauthorrefmark{1} Electrical engineering department, Information Technology University, Lahore 54000, Pakistan\\ \IEEEauthorrefmark{2} College of Science and Engineering, Hamad Bin Khalifa University (HBKU), Doha, Qatar \\
\IEEEauthorrefmark{3} Department of Computer Engineering, The University of Lahore, Lahore, 54000, Pakistan \\ 
\IEEEauthorrefmark{4}school of computer science and electronic engineering, University of Essex, UK\\
\IEEEauthorrefmark{5}James Watt School of Engineering, University of Glasgow, Glasgow, G12 8QQ, UK\\
\IEEEauthorrefmark{6}Artificial Intelligence Research Centre, Ajman University, Ajman, UAE\\
\IEEEauthorrefmark{1}\{mscs18011, mahboob.rahman\}@itu.edu.pk, \IEEEauthorrefmark{2}waman@hbku.edu.qa, \IEEEauthorrefmark{4}Qammer.Abbasi@glasgow.ac.uk
} 
\thanks{This work is supported in part by UK EPSRC under the grant: EP/X040518/1.}
}

\maketitle
\thispagestyle{plain}
\pagestyle{plain}

\begin{abstract}

Non-intrusive load monitoring (NILM) or energy disaggregation aims to extract the load profiles of individual consumer electronic appliances, given an aggregate load profile of the mains of a smart home. 
This work proposes a novel deep-learning and edge computing approach to solve the NILM problem and a few related problems as follows. 1) We build upon the reputed seq2-point convolutional neural network (CNN) model to come up with the proposed seq2-[3]-point CNN model to solve the (home) NILM problem and site-NILM problem (basically, NILM at a smaller scale). 2) We solve the related problem of appliance identification by building upon the state-of-the-art (pre-trained) 2D-CNN models, i.e., AlexNet, ResNet-18, and DenseNet-121, which are fine-tuned two custom datasets that consist of Wavelets and short-time Fourier transform (STFT)-based 2D electrical signatures of the appliances. 3) Finally, we do some basic qualitative inference about an individual appliance's health by comparing the power consumption of the same appliance across multiple homes. Low-frequency REDD dataset is used for all problems, except site-NILM where REFIT dataset has been used. As for the results, we achieve a maximum accuracy of 94.6\% for home-NILM, 81\% for site-NILM, and 88.9\% for appliance identification (with Resnet-based model). 




\end{abstract}



\section{Introduction}

Edge computing is a promising paradigm that could enable edge-intelligence, could reduce the latency between the sensing and actuation by closing the control loop within the consumer electronic devices, and is an enabler for federated learning on the consumer electronic devices \cite{edgeconsumer}. Thus, edge computing is anticipated to be an integral component of artificial intelligence (AI)-empowered consumer electronics devices in smart homes of the future \cite{edgeconsumersmarthome}. It goes without saying that edge computing will also empower the smart energy management/conservation systems, which are now considered as essential ingredients of smart green cities of future \cite{Kim2021}. This paper focuses on  non-intrusive load monitoring (NILM)---a building block of smart energy management/conservation systems---as one potential usecase scenario that could immensely benefit from edge computing. 


NILM, also known as energy disaggregation is an inverse problem whereby the goal is to estimate the power consumption profile of multiple individual consumer electronic appliances, given the aggregate power consumption profile of the mains of a smart home. As NILM could help identify the power usage patterns of individual appliances in a home, it could help realize novel energy conservation schemes for smart homes. It has been estimated that NILM could help reduce energy consumption up to 15\% in a household setting \cite{Fischer2008}. Thus, NILM is being boasted as one key enabler to realize smart homes and smart cities of the future. 

The idea of NILM at the appliance level was first conceived by G. W. Hart \cite{ghart1992nilm}. He modeled each appliance as a finite state machine, and his proposed NILM algorithm consisted of four major steps: data collection, event detection, feature extraction, and load identification. He argued that both steady-state analysis (i.e., fundamental frequency, harmonics, direct current) and transients analysis (shape, size, duration of transients) provide valuable information for NILM purpose. 


NILM framework has spawned many other interesting problems. For example, NILM finds its application in future smart grid systems where power utility companies will charge a customized tariff to their clients, based upon the appliance-level power usage data collected from their homes via smart meters \cite{myhmmpaper}. NILM could also help consumers to cut down their energy bills by providing them a detailed picture of the power usage report of appliances in their homes. Some other examples include: energy audit of buildings, estimating the remaining useful life of an appliance, inferring the build quality of an appliance, demand response prediction, inferring the lifestyle of consumers in a home (based upon power consumption behavior of appliances in a household) etc\footnote{The interested reader is referred to the survey/review articles \cite{huber2021review}, \cite{angelis2022nilm}, \cite{revuelta2017non}, \cite{ruano2019nilm} for a more systematic and detailed review of the works on NILM as well as the existing NILM datasets.}.

\subsection{Contributions}

The key contributions of this work are as follows:

\begin{itemize}
  \item We present a novel sequence-to-sequence deep learning (DL) model---so-called seq2-[3]point model, which efficiently predicts the power consumption of individual appliances from power consumption data of the aggregate signal (of mains of a home, or of a site).
  \item We do appliance signature classification by customizing the state-of-the-art 2D convolutional neural networks (CNN), i.e., AlexNet, ResNet18, DenseNet-121. To this end, we translate the 1D time-series data (of aggregate power of the mains and individual appliances) into 2D (image) data using two different transforms: Wavelet transform, Short-time Fourier transform (STFT), and their fusion (Wavelets + STFT).
  \item We do qualitative inference about an appliance (i.e. the state it is in, its health and well-being) by analysing its power consumption patterns.
\end{itemize}

Note that both proposed DL-based methods for NILM and appliance classification utilize the {\it transfer learning} (TL) approach where the weights of the convolutional layers are frozen after training for one appliance is complete, and are reused to learn the signatures of other appliances. This greatly reduces the computational requirements of the two proposed DL-based methods, and makes them amenable for edge computing (by the individual consumer electronic devices). Last but not the least, low-frequency REDD dataset was used for all the problems, except the site-NILM task where REFIT dataset was used.

\subsection{Outline}
Section II summarizes the selected related work. Section III describes the essentials details of the two datasets used, as well as critical data pre-processing steps performed on both datasets. Section IV outlines the details of our proposed seq2-[3]point DL \& TL-based method for home-NILM and site-NILM. Section V presents our proposed DL \& TL-based method for appliance identification. Section VI does some qualitative inference from the power consumption analysis of appliances. Section VII provided some selected results. Section VIII concludes the paper.

\section{Related Work}

NILM and its derivative problems have consistently attracted quite some attention during the past two decades. Further, with the rise of sophisticated machine/deep learning techniques during the last few years, many researchers have proposed efficient algorithms that learn/train on huge datasets with the ultimate aim to exhibit good performance when tested on unseen data (with potentially different distribution than the data the algorithm was originally trained on). Given the large volume of works on NILM, we discuss only selected related work (most relevant to our work), which is classified into three broad categories: classical signal processing methods, machine/deep learning methods, and NILM prototypes. 


{\it Signal processing methods:} 
In \cite{kolter2011redd}, authors present a new NILM dataset called REDD dataset, and utilize a variant of hidden Markov model (HMM) known as factorial HMM to model the NILM problem. They train the factorial HMM by means of Baum-Welch expectation-maximization algorithm, and eventually estimate the actual (hidden) state of each appliance using the observations of the aggregate signal from the whole-home power monitor. \cite{kolter2010energy} solves the energy disaggregation problem by using a variant of sparse coding known as discriminative sparse coding, along with structured prediction methods. The data used in this work was taken from a European company {\it Plugwise} who collected it from 590 homes (and 10,165 unique devices), over a period of two years. \cite{bouhouras2019nilm} utilizes a custom equipment to collect data in a residential setting, and proposes to construct the unique load signature for each appliance using the first three odd harmonic current vectors. \cite{machlev2018modified} formulates the NILM problem as a constrained optimization program that is based on a combinatorial (cross-entropy) method, and solves it using a new penalty-based method that utilizes Hamming distance between the current and previously known states of the appliances. They showcase the performance of their proposed method on publicly available REDD and AMPds datasets, \cite{kolter2011redd} and \cite{makonin2013ampds}.

{\it Machine learning approaches:}
\cite{figueiredo2012home} collects the custom data, pre-processes it, and utilizes support vector machine (SVM) and 5-nearest neighbors (NN) algorithms to determine the switching time (i.e., the time instant when an On/Off transition occurs) for each appliance, which in turn helps in identification of the unique load signature of each appliance (and hence the NILM). \cite{figueiredo2011experimental} exploits the step-changes in the features (active power, reactive power, power factor) of individual appliances when they switch on or off, in order to extract the appliance-level power consumption from the aggregate power signal. Specifically, authors utilize SVM model and k-NN methods, and report an accuracy of 90\% and 98\%, respectively. \cite{lin2011applications} utilizes a two-stage fuzzy classifier for load identification that is based upon the transients information extracted from the aggregate power signal. They first learn the coarse parameters of their proposed classifier by means of the fuzzy C-Means clustering algorithms, and later refine the coarse parameter estimates by means of the two algorithms: error back propagataion and genetic algorithm. \cite{puente2020non} proposes an unsupervised learning approach that builds upon soft computing methods (i.e., fuzzy clustering), and tests the performance of their proposed methods on a controlled dataset (collected via smart meters of various households) and UK-DALE dataset \cite{kelly2015uk}.

{\it Deep learning approaches:}
\cite{kim2017nonintrusive} proposes a long short-term memory (LSTM)-based recurrent neural network (RNN) model to solve the NILM problem for a peculiar scenario where many appliances could have a very similar load profile; furthermore, each appliance could be in one of many states at any given time. Authors in \cite{kelly2015neural} train three different deep learning models, i.e., an RNN , a denoising auto-encoder and a regression based model, on a custom dataset to perform the NILM task. \cite{sadeghianpourhamami2017comprehensive}, on the other hand, utilizes a random forest classifier to do feature selection for NILM, whereby the candidate features are eliminated one by one until a few most relevant features remain are left in the end. Further, \cite{makonin2015nonintrusive} argues that there is a lack of a robust performance metric to reliably assess the performance of various competing NILM solutions, proposes one such metric and tests its performance using the AMPds dataset \cite{makonin2013ampds}. \cite{ciancetta2020new} utilizes a dataset called BLUED, computes the spectrogram of the energy data, feeds it to a convolutional neural network (CNN), that returns the load signature of each appliance as well as its current state (On or Off). This work reports an F1 score and an accuracy of 99.8\% and 87.9\% respectively. \cite{gomes2020pb} utilizes the REFIT dataset, and proposes the pinball (or, quantile) loss function instead of more common mean squared error (MSE) loss function on a custum-built deep neural network, due to the fact that the MSE function tends to guide the model towards the median of the distribution, which may not be desired because NILM tends to penalize large errors.    





Up to this point, all the deep learning-based works discussed above consider a single-task learning approach. That is, each of the above-mentioned works trains a neural network exclusively for each appliance. However, in reality, more than one appliance could be simultaneously active at any given time, thus the inter-dependency between various appliances needs to be incorporated as well. Such methods called multi-task deep learning methods are discussed next.


{\it Multi-task deep learning methods:}
The study \cite{faustine2020unet} takes this limitation into account and proposes a 1D CNN model based on popular U-Net architecture, which exploits multi-label learning and multi-target quantile regression, in order to learn the power usage and exact state of each of the many appliances (from the UK-DALE dataset). The work in \cite{nalmpantis2020time} constructs a number of time-domain and frequency-domain sequences from the aggregate power signal, implements decision trees and feedforward neural network classifiers, and test the performance of their proposed approach on two datasets: UK-DALE and REDD. 
Authors in \cite{machlev2020dimension} propose the use of principal component analysis (PCA) as an effective method for dimensionality reduction (of aggregate power data) with minimal information loss, and showcase its performance for two NILM datasets, AMPds dataset and a private dataset. 
The work \cite{de2017bayesian} considers the BLUED dataset \cite{filip2011blued}, and proposes two event-detection methods: chi-squared goodness of fit test, and cepstrum smoothing. Both models are then fine-tuned using the surrogate-based optimization (SBO) method. \cite{faustine2020multi} utilizes the PLAID dataset, and makes use of Fryze power theory to split the activation current into active and non-active components, which are then transformed into image-like representations using a Euclidean distance similarity function, and are fed into a multi-label CNN classifier for appliance identification. \cite{garcia2020fully} considers the aggregate load profile data collected from a non-residential setup (basically a hospital building). They propose two deep learning architectures, i.e., a CNN-based denoising auto-encoder (dAE) architecture and a vanilla dAE, and test its performance on their custom collected data. \cite{do2016applications} implements a range of deep learning architectures, i.e., RNNs (both LSTM and GRU), CNNs (RCNN and residual), and tests them all on REDD dataset in order to do appliance classification as well as appliance power consumption estimation.  \cite{chen2021temporal} constructs various kinds of temporal and spectral signatures from the measured data of aggregate load and feeds them to a two-stream CNN (built upon AlexNet) which does the feature extraction and performs load classification. The work \cite{kukunuri2020edgenilm} focuses on the design of a light-weight neural network (based upon seq2point CNN) suitable for edge-computing. To this end, they propose a multi-task learning-based architecture, do filter pruning and neuron pruning to lower the memory and computational costs, and test its accuracy on REDD and UK-DALE datasets.

{\it NILM prototypes:}
The work \cite{ahmed2020edge} proposes MobileNet, an edge-computing framework that provides a reliable solution to {\it NILM as a service} problem. There, authors implement and test a light-weight neural network solution for NILM, and then compress their model further using TensorflowLite so that the end users could access the NILM results in real-time on an app in their smartphones. \cite{biansoongnern2016nonintrusive} measures the aggregate steady-state active power and reactive power consumed by a group of four appliances using a custom equipment and feeds the collected data to an artificial neural network (ANN) in order to train it for NILM of the four appliances. \cite{biansoongnern2016non} builds and tests an NILM prototype that collects aggregate power data, detects transients, separates the power consumption of two appliances (a fridge and an air conditioner), and displays the results on a web/GUI interface. 


 \section{NILM Datasets and Data Preprocessing }

This section first describes the essentials details of the two public NILM datasets (REDD and REFIT) utilized in this work\footnote{For the interested, other than REDD and REFIT, there are a few more public NILM datasets too, e.g., BLUED, PLAID, AMPds, UK-DALE. }. We then outline the key data preprocessing steps that were carried out on the two datasets in order to prepare them for the training of the various proposed deep learning models.

\subsection{Datasets used in this work}
\subsubsection{REDD Dataset \cite{kolter2011redd} }
The REDD dataset consists of 119 days of data collected from 10 homes. For each home, the data was collected from the mains and (up to 20) plug-level monitors, and is split into three categories: low-frequency data, high-frequency data, and high-frequency raw data.
For each home, the low-frequency data consists of a time series of the aggregate power consumed by the home (at a rate of 0.3 Hz) as well as the time series of the power consumed by each of the many appliances in the home (at a rate of 1 Hz). 
The high-frequency data, on the other hand, consists of the raw current and voltage waveforms for the mains as well as each of the appliances (at a rate of 15 KHz), along with the corresponding UTC timestamps.

\subsubsection{REFIT Dataset \cite{278e1df91d22494f9be2adfca2559f92} }
The REFIT dataset contains both aggregate and appliance-level data collected after every 8 seconds, from 20 homes over a period of two years. Unlike REDD, REFIT dataset provides additional information about the various composite sites in a house. For example, load profile of a "Computer site" implies the aggregate power usage of two or more appliances, e.g., desktop computer, laptop, charging stations, printer etc. Metadata about the homes is also present;
it includes the following: number of occupants in a home, construction year, size of the home, number of appliances etc. 


\subsection{Pre-processing on REDD dataset}

Initial investigation of REDD dataset revealed that the task of synchronizing the time series of the mains with the time series of various appliances for each home is a non-trivial task, due to the presence of missing data. This may be due to loss in power, a malfunctioning equipment, and network problems. Furthermore, in REDD dataset, the mains readings are separated by an interval of either 3 or 4 seconds, whereas appliance readings are separated by 1 second. But, the time synchronization task requires that the timestamp of the appliance needs to be matched with the mains timestamp, for each data point. 

{\it Synchronization of mains time-series with the time-series of appliances:}
The synchronization challenge prompted us to redesign/re-arrange the dataset such that each house gets exactly one file that contains the power readings of the mains and all the appliances, along with a file that contains the appliance labels.
Let us elaborate this further by taking House 1 as an example. For House 1, A new data file was created with a total of $N+3$ columns, where $N$ represents the number of appliances in House 1. Then, the first 3 columns of the new data file store the UTC timestamps corresponding to the data points, and the readings of mains 1 and mains 2, respectively. The (reference) UTC timestamps were taken from appliance-level data as timestamps for the appliance-level data were all in harmony with each other. Thereafter, all those readings of mains 1 and mains 2 were discarded whose timestamps were not in harmony with the reference UTC timestamps. The remaining $N$ columns store the power readings of the $N$ appliances. For those UTC stamps where mains data was not available, 0 was placed.

{\it Final Data Files for home-NILM: } \label{finaldatafiles}
The final step was to create a custom dataset to feed it to the proposed deep learning model. Because the model takes data of 1 appliance at a time, each appliance has a test and train csv file containing of 2 columns, where 1\textsuperscript{st} column represents the aggregate signal, while the 2\textsuperscript{nd} column represents appliance data. The aggregate signal was constructed by adding the power readings of mains 1 and mains 2 together. 

{\it Data Normalization: }
Data for each appliance as well as the aggregate signal was normalized using the standard normalizing technique: $Z = \frac{X-\mu}{\sigma}$, where \textbf{\(\mu\)}, \textbf{\(\sigma\)}, $X$, $Z$ represent the mean of the data, the standard deviation of the data, the un-normalized data, and normalized data, respectively.

\subsection{Pre-processing on REFIT dataset}
As the reader may recall, we utilize the REFIT dataset in order to do site-NILM---the home-NILM problem toned down to a smaller scale of a {\it site}. As an example, consider a {\it computer site}. The power readings collected from a plug-monitor connected to a computer site will encapsulate the power usage by multiple appliances, e.g., desktop computer, monitors, a speaker system, printer, router etc. Then, the natural task of site-NILM is to estimate appliance-level power usage from the site-level aggregate power data. To this end, we consider a computer site that consists of a desktop computer and a monitor\footnote{We considered only two appliances for site-NILM because they were the only appliances that were common among all the houses.}. Then, for site-NILM, we construct a training (test) dataset out of two (one) houses with a computer site as follows. By applying different thresholds on the aggregate power $X$ drawn, the computer site could be classified into being in one of the four classes (states) A, B, C, D. That is, if $X<10$ Watts, then computer site is in class (state) A: monitor off and computer in sleep. $10<X<15$ Watts leads to class B: both monitor and computer in sleep. $15<X<80$ Watts leads to class C: monitor in use but computer in sleep. Finally, $X>80$ Watts leads to class D: both monitor and computer in use.
Eventually, the training and test dataset have 3 columns each: 1\textsuperscript{st} column contains the aggregate signal readings, 2\textsuperscript{nd} column contains the appliance readings, and 3\textsuperscript{rd} column contains the class labels. 

\section{Home-NILM \& Site-NILM}

This section succinctly describes the proposed seq2-[3]point DL method which we utilize to solve the home-NILM problem and the site-NILM problem. 

\subsection{Home-NILM}

{\it Seq2-point CNN model.}
This work builds upon the reputed seq2-point CNN model of \cite{d2019transfer} in order to do NILM. A brief description of the seq2-point model is as follows. Time synchronize the load profile of an appliance with that of the mains, consider a specific time duration to determine the window size, and fetch synchronized data vectors/windows for both the mains and the appliance from their load profiles. Then, the seq2-point CNN model takes at its input a window/vector of $L$ data points of the aggregate mains power signal, and returns a scalar at its output, which is by definition the mid-point of the corresponding appliance window/vector. 

{\it Proposed Seq2-[3]point CNN model.}
We customize the seq2-point CNN model such that the model takes as its input a window/vector of $L$ data points of the aggregate mains power signal as before, but now returns a vector of three elements at its output. The elements of output vector are the first point, mid-point, and last point of the corresponding appliance window/vector. Another notable difference is that the proposed seq2-[3]point model has its own (two) fully-connected (FC) layers. Thus, in the proposed seq2-[3]point model, there are a total of five convolutional (conv.) layers followed by a flatten layer, followed by two custom FC layers. Table \ref{tab:convlayers} describes the attributes of each of the five conv. layers $C_i$.

\begin{table}[ht]
\centering
  \caption[Conv. layer details for NILM CNN]{}
  \small {Attributes of the conv. layers of the proposed seq2-[3]point CNN}
  \begin{center}
    \begin{tabular}{c|c|c|c|c|c}
      \hline {\(C_i\)} & {\(in_c\)} & {\(out_c\)} & {\(k_s\)} & $S_t$ & $P$ \\ \hline \hline
      C1 & 1 & 30 & 10 & 1 & 0 \\ \hline
      C2 & 30 & 30 & 8 & 1 & 0 \\ \hline 
      C3 & 30 & 40 & 6 & 1 & 0 \\ \hline 
      C4 & 40 & 50 & 5 & 1 & 0 \\ \hline
      C5 & 50 & 50 & 5 & 1 & 0
    \end{tabular}
  \end{center}
  \label{tab:convlayers}
\end{table}

\begin{figure}
    \centering
    \subfigure[]{\includegraphics[width=0.48\textwidth]{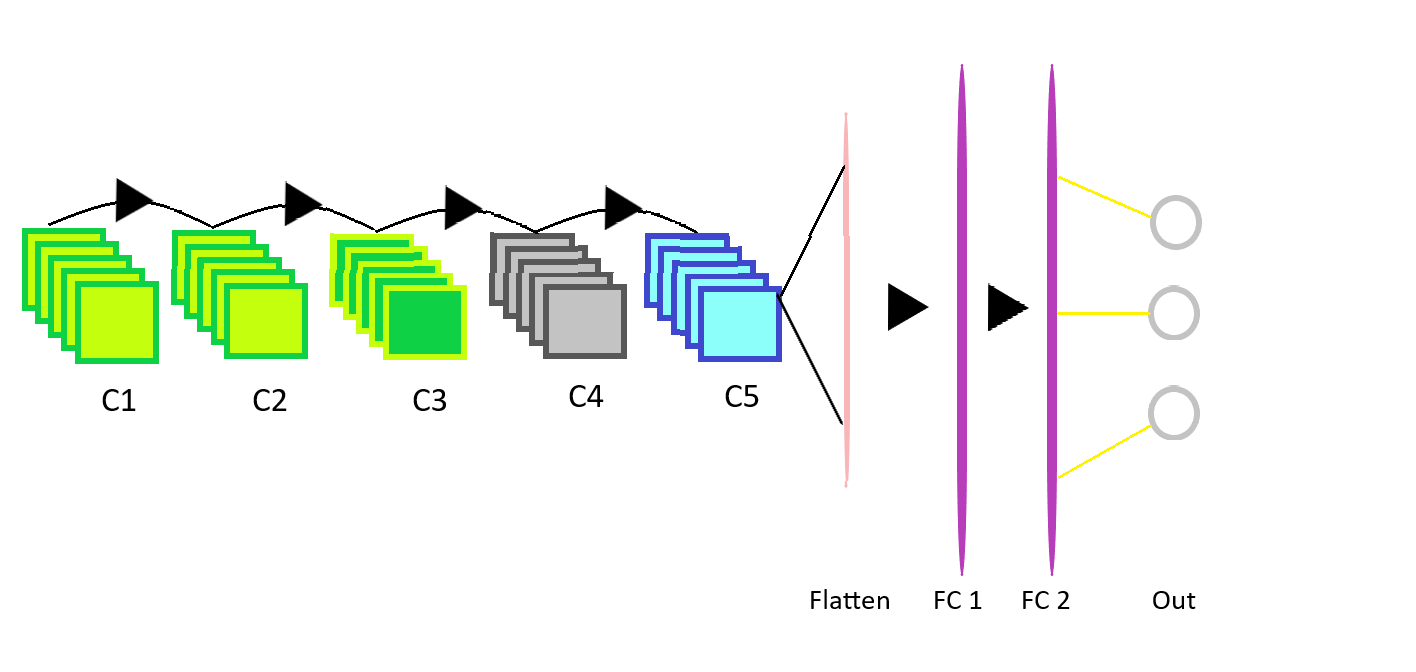}} 
    \caption{Architecture of the proposed seq2-[3]point CNN model. It consists of five convolutional layers, two custom FC layers, and outputs a sequence/vector of size three.}
    \label{fig:ourcnn}
\end{figure}

Here, $k_s$, $S_t$, $P$ represent the filter size, stride and padding, respectively. 
Note that each conv. layer has a ReLu activation applied.
A total of {48,550} parameters are returned by the final conv. layer. These parameters are then flattened, and passed on to the FC layers of the model. The first FC layer takes as input 48,550 parameters and outputs 1300 parameters, and uses the ReLu activation function. The second FC layer takes 1300 parameters and returns 3 parameters (i.e., start point, mid-point and end point of the appliance window).  
Once the very first appliance was learned by the proposed model, the convolution layers were freezed for the sake of {\it transfer learning} so that the rest of the appliances so that each appliance has similar conv. features. Fig. \ref{fig:ourcnn} shows in detail the architecture of the proposed seq2-[3]point CNN model.


{\it Train Test data construction.} 
Data used is the pre-processed data obtained in Section III-B. 
Further, let {\(S\)} and {\(E\)} be start and end index, {\(L\)} be length of a window, {\(B\)} be the batch size. Then, we begin as follows: $S=0, E=L=1000, B=20000, offset=35$. 
Next, update {\(S\)} and {\(E\)} as follows: $S = S + offset$ and $E = E + offset$.
This process is repeated until a total samples of {\(B\)} is reached. Test data is constructed using the same method. 

{\it Model Evaluation.}
The input to the model is a window containing 1000 points, with a batch size of 20,000. Thus, Train X has a shape of {\((20000, 1000)\)}, formally described as {\((B, L)\)}. The output returned by conv. layer $C_1$ is (1000, 30, 991, 1). The output returned by conv. layer $C_2$ is (1000, 30, 984, 1). The output returned by conv. layer $C_3$ is (1000, 40, 979, 1). The output returned by conv. layer $C_4$ is (1000, 50, 971, 1).
The final conv. layer, when flattened gives a total of 48,550 trainable parameters which are passed to the FC layers of the model. Size of each convolution layer can be computed as follows: $C_{Out} = \frac{W - k_s + 2P}{S_t} + 1$, where {\(W = \)} width of the matrix. This work uses mean squared error (MSE) as criterion, Adam optimizer, and a learning rate of 0.001. 
\\
{\it Back propagation.} 
The back pass done by the model in this work required gradients for the conv. layers for the very first appliance. Therefore, during the backward pass, the weights are updated as follows: $W_n = W_o - \alpha \times \frac{\partial e}{\partial W_o}$, where {\(W_n\)} is the new weight, {\(W_o\)} is the old weight, {\(\alpha\)} is the learning rate and {\(e\)} is the error from the previous layer. For the rest of the appliances, weights for the very first trained appliance were used as a base ({\it inline with the transfer learning principle}). Thus, the gradients for the conv. layers were freezed so that each appliance has similar conv. features. 

{\it Translation of MSE to accuracy.}
Because the model learns to predict power values for a given appliance from the aggregate signal, the model evaluation process requires caution. In this work, we take the absolute difference $D$ between the model predictions $PD$ and the actual ground truth $GT$ as follows: $D = |PD - GT|$. We then compare $D$ against a threshold $\tau$. If $D < \tau$, the predicted value is considered a correct predicted value, and vice versa. We carefully define different threshold values for different appliances. An important note. Model evaluation was done in batches, so {\(PD\)} and {\(GT\)} are essentially arrays, which makes {\(D\)} an array as well.


\subsection{Site-NILM}
We utilized the same CNN architecture as described in section IV-A for site-NILM. As the predictions were to be classified in 4 different classes (see Section III-C), training the data as is did not help the model converge. So, we came up with a solution of first normalizing the data, and during the testing phase, denormalized the data again to be able to assign classes. For normalising purposes, Gaussian normalization was used. During the test phase, the denormalizing process is given by the following: $X_o = (X_n * \sigma) + \mu$, where \textbf{$X_o$} is the original value, \textbf{$X_n$} is the normalized value, \textbf{{\(\sigma\)}} is the standard deviation and \textbf{{\(\mu\)}} is the mean.


\section{Appliance identification}

This section outlines the pertinent details of our proposed method for appliance identification whereby we utilize the appliance-level load signatures from the low-frequency data of the REDD dataset. Since the proposed custom CNN model is to be trained on 2D images, we first discuss our methodology to convert the 1D time-series into 2D images. Subsequently, we describe the key details of the proposed CNN model. 

\subsection{Custom 2D dataset generation}

We apply two kinds of transforms on the REDD low-frequency data: i) Wavelet Transform, ii) Short Time Fourier Transform (STFT), in order to translate the 1D dataset (load profiles of appliances) into a 2D dataset (consisting of Wavelet and STFT images). Below, we describe both transform operations, one by one.

\subsubsection{Wavelet Transform}

The load profile data of each appliance was projected onto a 2D-space by passing it through a Wavelet function which returned the detailed and approximation coefficients. Precisely speaking, the load profile data of each appliance was split into windows/frames of duration one second, and a sliding window mechanism (with an offset of 0.5 seconds) was used in order to generate enough Wavelet images. The mother wavelet used was Mexican hat, and scale range was varied in the range 1-500. 
{Algorithm~\ref{wt-algorithm}} shows a generalised version of wavelet generation algorithm. 

\begin{algorithm}[thb]
\caption{Wavelets-based image generation algorithm}
\label{wt-algorithm}
\begin{algorithmic}
  \REQUIRE $H_{i}$: House Number, $C_{i}$: Channel Number, $Max_{r}$: Maximum data point length, $It_{m}$: Maximum Iterations, $D_{o}$: Data Offset point
  \ENSURE $S$: Wavelet coefficient Spectrogram 
  \STATE $S_{p} \gets 0$   
  \STATE $E_{p} \gets Max_{r}$   

  \STATE $PR \gets \textsc{Get-APPLIANCE-READINGS}(H_{i}, C_{i})$
  \FOR{$i=0,It_{m}$}
    \STATE $ODD \gets PR[S_{p} : E_{p}]$
    \STATE $SG \gets \textsc{Generate-Wavelet-Spectrogram}(ODD)$
    \STATE $ODD \gets \emptyset$
    \STATE $S_{p} \gets S_{p} + D_{o}$
    \STATE $E_{p} \gets S_{p} + Max_{r}$
  \ENDFOR
 \end{algorithmic}
\end{algorithm}

The Wavelet-based dataset was generated using the Python library {PYWT}, also known as py-wavelets. 
The final classes were not divided into houses because the goal was to classify any appliance based on this signature. There are a total of 19 unique appliances from all 6 houses combined, excluding the mains. {Figure~\ref{fig:2}} (first column from left) shows load profiles of three appliances (i.e., refrigerator, microwave, kitchen outlets), while {Figure~\ref{fig:2}} (third column from left) shows the corresponding 2D image consisting of wavelet coefficients (here, x-axis represents the time, while the y-axis represents the signal frequencies).


{\it Data Augmentation:}
Image augmentation was applied after the images were generated because for some appliances, there wasn't simply enough data to go with. For example, some appliances have data readings which exceed 1 Million, whereas some appliances struggle to cross the 300K barrier. Due to this reason, images in each classes were subjected to augmentation, in order to increase the dataset size artificially. A range of augmentation techniques was applied on the Wavelets dataset, e.g., rotation by a certain angle in degrees, shearing image, cropping image etc. The benefit of the augmentation was that for each train-test data split, each class had an equal number of images in each of the split. \\

\subsubsection{Short Time Fourier Transform}

For the computation of STFT spectrograms, {Algorithm~\ref{wt-algorithm}} was used again, but the data points were passed to an STFT function (instead of the Wavelet function).
Again, the classes were not divided into houses because the main goal was to classify appliances based on their unique signatures. {Figure~\ref{fig:2}} (second column from left) shows the STFT spectrogram of the three appliances (here, x-axis represents the time, while the y-axis represents the frequency (in Hz). As in the case of Wavelet transform, the data was simply not enough for some appliances. Thus, data augmentation was applied again in order to make sure that train and test classes had an equal number of images. The main purpose for computing STFTs was to generate a fused electrical signal by combining two different electrical signals, i.e., the Wavelet transforms and STFTs (in our case).


\subsubsection{Fused Spectrograms: Wavelet + STFT}
\begin{figure*}
    \centering
    \subfigure[]{\includegraphics[width=0.98\textwidth]{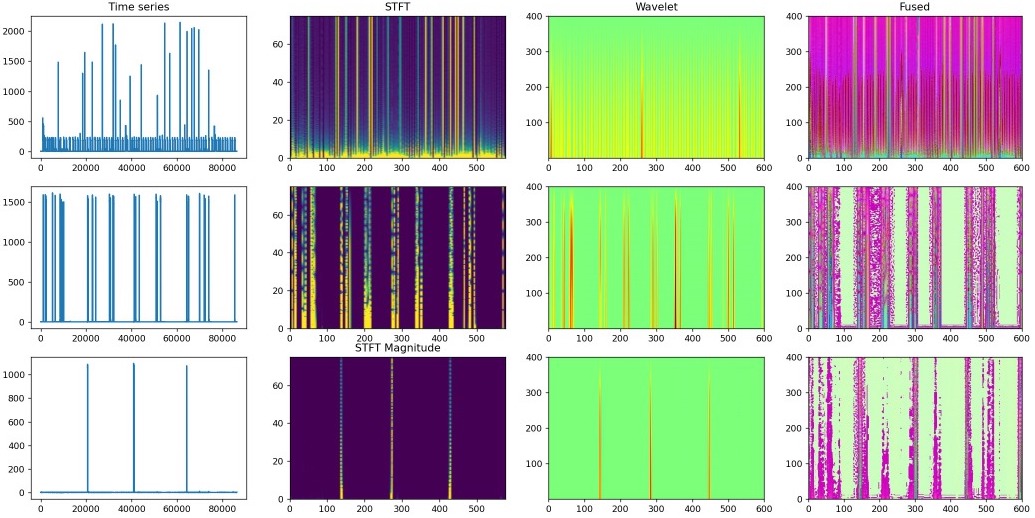}}
    \caption{ Starting from left, first column shows (time-domain) load profile of three appliances (refrigerator, microwave, kitchen outlet); second column shows STFT spectrograms of the three appliances; third column shows the Wavelet of the three appliances; fourth column shows the fusion of two transforms (Wavelet+STFT) for the three appliances. Overall, it can be appreciated that Wavelets and STFT for three appliances are distinct, and thus, could help in appliance identification. }
    \label{fig:2}
\end{figure*}

As the name suggests, the third type of electrical signature of an appliance this work proposes is obtained by combining Wavelet transform and STFT images (basically, adding the two images, pixel-by-pixel). Data augmentation was done for this dataset too, but only after the dataset was generated using the original Wavelet transform and STFT images. {Figure~\ref{fig:2}} (right-most column) shows
the fusion of both transforms for the three appliances.



{\it Train-Test Split of the custom 2D Dataset: }
With 19 classes in total for each of the transforms (Wavelet, STFT, Fused), the train-test split was done as follows. Wavelet Transform: Total Train images: 600, Total Test images: 200. Wavelet + STFT: Total Train images: 600, Total Test images: 200.
Note that about 75\% to 80\% of the images in each category were the originals, the rest were augmented images.

\subsection{Custom CNN model design for Appliance Classification}

\subsubsection{Proposed CNN models for appliance classification on Wavelet-based dataset }
The wavelet signature of appliances were classified using a simple deep neural network (NN) and multiple CNNs. The simple deep NN was the maiden network used in this study for classification, and its details are as follows:
Input size: 56 x 34, Total no. of hidden layers: 2, Hidden layer 1 size: 500, Hidden layer 2 size: 150, Output layer size: 20, Loss function: Cross entropy, Optimizer: SGD.

Two state-of-the-art pre-trained 2D-CNNs, namely ResNet18\cite{he2016deep} and AlexNet\cite{krizhevsky2012imagenet} were used with the modification that for both of the networks, the conv. layers were preserved, but the FC layers were cut-out and replaced with custom FC layers. Both models used cross-entropy loss function and SGD optimizer during training. 

For Resnet18, custom FC layers were added as follows: FC1 = [2500,2000], and FC2 = [2000, 1500], FC3 = [1500,500], and FC4 = [500, 20].

For Alexnet, custom FC layers were added as follows: FC1 = [4096,1024], and FC2 = [1024, 20].

\subsubsection{Proposed CNN model for appliance classification on Wavelet+STFT-based dataset }
The fused signatures of appliances were classified using a state-of-the-art CNN, called Densenet-121 \cite{huang2017densely}. 
As before, the conv. layers were not changed but the FC layers were completely stripped off, and custom FC layers were added as follows: FC1 = [1024,512], and FC2 = [512, 20]\footnote{This configuration of FC layers was the best this study could come up with, with main reason being GPU running out of CUDA memory.}. Some Additional details are as follows: FC1 activation function: Relu, FC2 activation function: Softmax, Loss function: Cross entropy, Optimizer: SGD.

\section{Inference about the appliance behaviour}

In addition to (home and site) NILM and appliance identification, we also study the related problem of appliance usage pattern and behavior analysis using the REDD dataset. To this end, we do a high-level performance comparison of the refrigerators---the most common appliance in the REDD dataset (as 5 out of 6 houses have refrigerators). Specifically, we extract from the dataset the maximum and average power consumption of the refrigerators over a period of two days. Additionally, we define following five classes/states for each refrigerator based upon the amount of the transient power $X$ drawn: stable/steady state ($X<3$), minor increase or decrease in power drawn ($3<X<10$), sudden large increase or decrease in power drawn ($10<X<50$). This is done by computing the pairwise difference of consecutive elements, given a list of appliance power readings.

\begin{figure}
    \centering
    \subfigure[]{\includegraphics[width=0.24\textwidth]{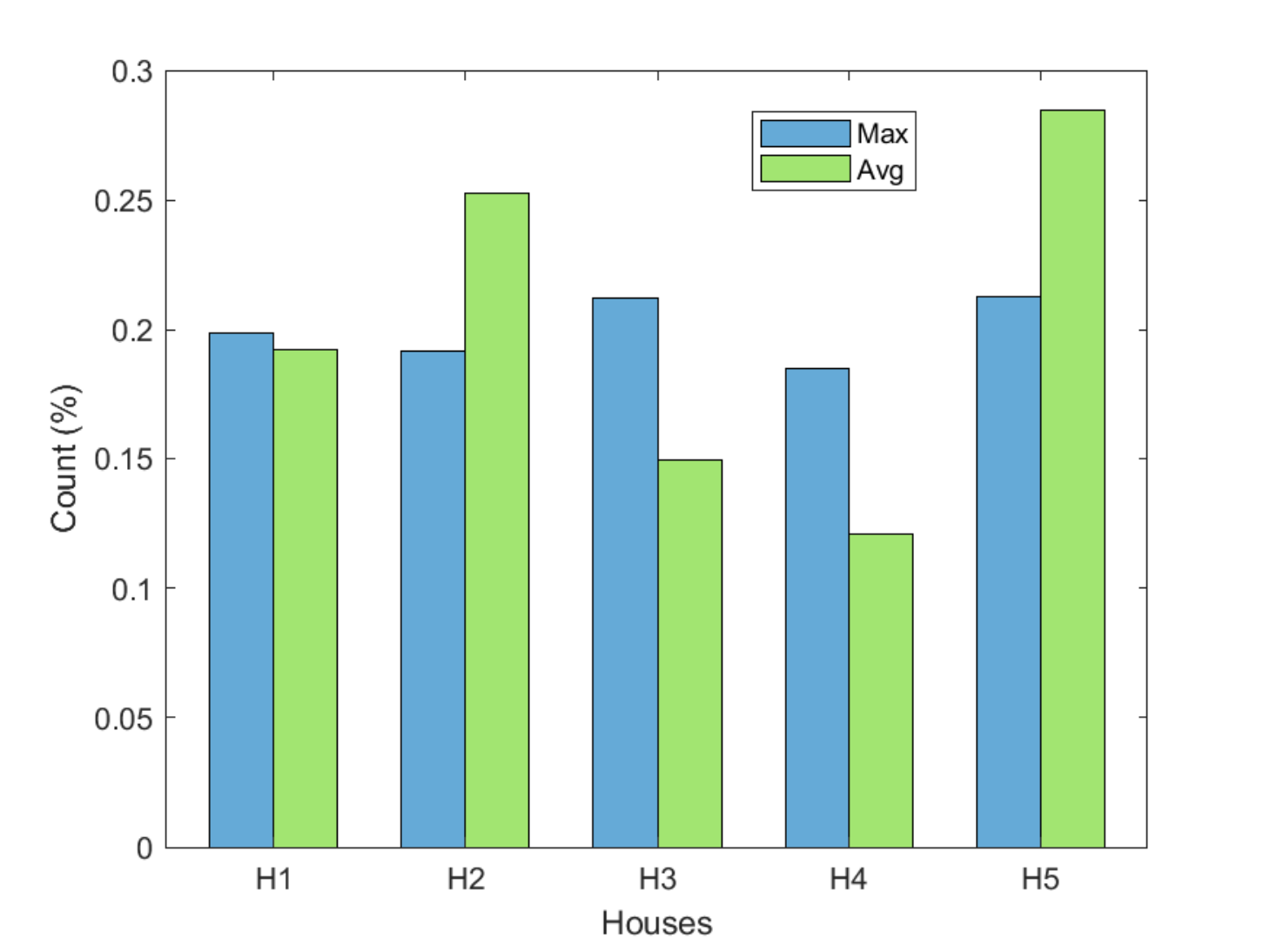}} 
    \subfigure[]{\includegraphics[width=0.24\textwidth]{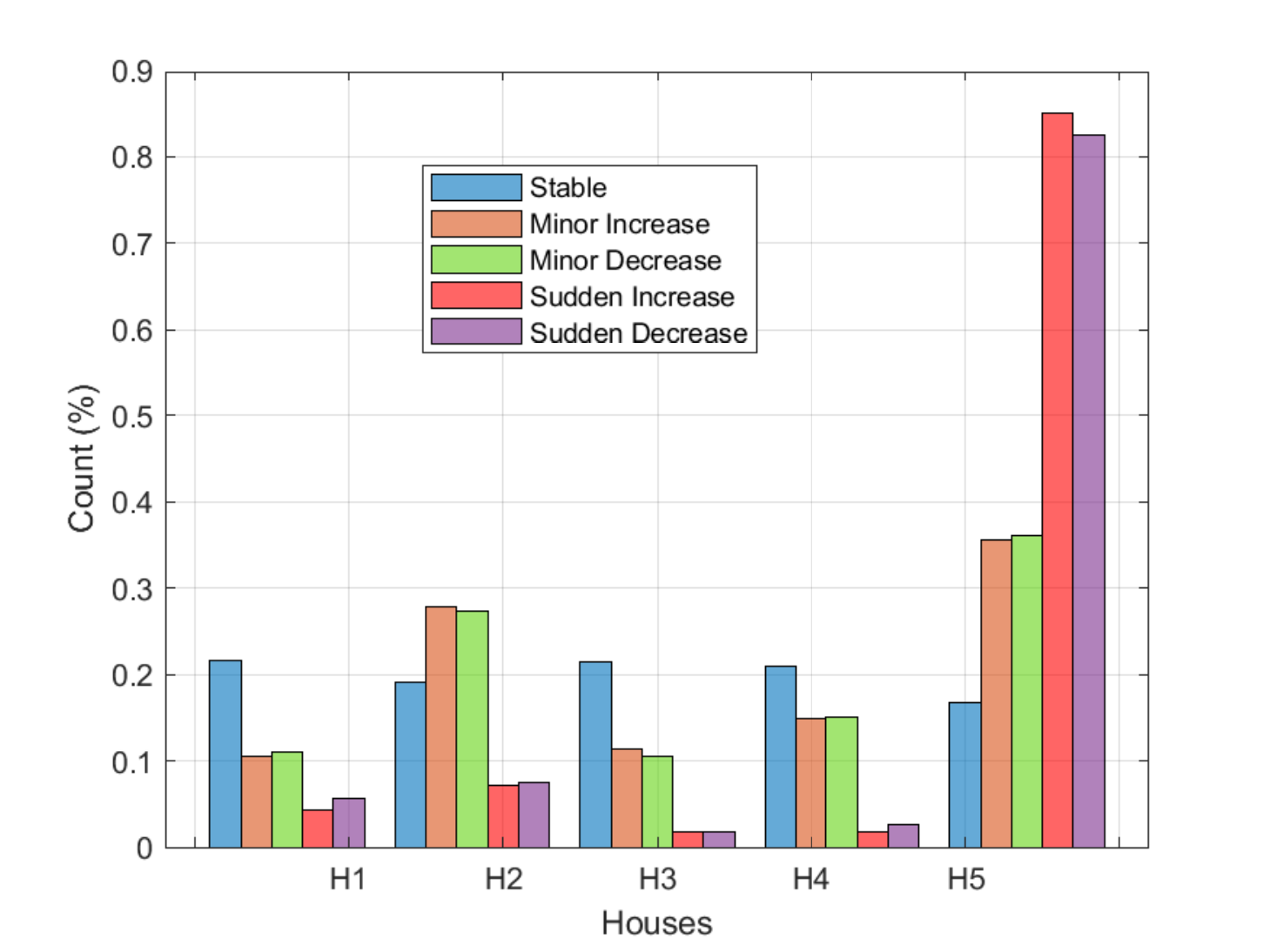}} 
    \caption{The histograms of (refrigerators across five homes): (a) Max and avg. power consumption (b) five power states. }
    \label{fig:minmaxpower}
\end{figure}

Fig. \ref{fig:minmaxpower} (a) shows that there is not much (relative) disparity in the max. and average power usage by the refrigerators in five homes (over a period of two days). This in turn indicates that even though the refrigerators in five homes may potentially have a different model, manufacturer and year of making, they seem to be in similar health condition, have a similar build quality, and undergo similar kind of use by their consumers. 

Fig. \ref{fig:minmaxpower} (b) indicates that the sudden large changes in power drawn are most prominent in the refrigerator of Home 6. This may imply any of the following: appliance overload, open door problem, failure of a component (temperature sensor, regulator, compressor etc.), appliance approaching its end-of-life. 

\section{Experimental Results}

\subsection{Computing Machine Specs.}
The training and testing of the proposed deep learning models on the REDD and REFIT datasets was done on an HP laptop machine with the following specifications: model Omen 15, Intel Core i7-7700HQ (KabyLake Architecture) CPU, Nvidia 1050ti 4GB GPU, 16 GB RAM, 1 TB HDD. 


\subsection{Evaluation Metrics}
We use accuracy and F1 score as the main performance evaluation metrics. Since accuracy is most meaningful for a balanced dataset, we minimized the class imbalance by means of data augmentation. 

\subsubsection{Accuracy}
The accuracy of a dataset is the ratio of number of correct predictions $Samples_c$ by the model and total number of samples $Samples_t$: $Accuracy = \frac{Samples_c}{Samples_t} \times 100 $. 

\subsubsection{F1 Score}
F1 score is a statistical measure to rate a model's performance. It is the harmonic mean of 2 factors, precision and recall. Precision is obtained by dividing total number of correctly classified elements, i.e. True Positives by total positively classified elements, i.e. True Positives (TP) + False Positives (FP). Thus, $Precision = \frac{TP}{TP + FP}$. Recall is obtained by dividing total number of positively classified samples by the total number of samples that should had been marked as positive, i.e., True Positives and True Positives + False Negatives (FN) respectively. Thus, $Recall = \frac{TP}{TP + FN} $.
When precision and recall in hand, F1 score is calculated as: $F1 = 2 * \frac{P*R}{P+R}$, where $P$ is precision and $R$ is recall.
 
\subsection{Home-NILM \& site-NILM Results}

\subsubsection{Home-NILM results}
For home-NILM, the proposed seq2-[3]point CNN was trained for four appliances. Table {\ref{tab:nilmmetrics}} showcases the test accuracy obtained by the proposed method for all the four appliances. Note that to compute the accuracy, the predictions provided by the proposed DL model were translated to the notion of accuracy, as explained in section IV-A (see last paragraph). Table {\ref{tab:nilmmetrics}} demonstrates that the accuracy of the proposed DL model remained quite high, and varied from 86.58\% to 94.6\%, for all the four appliances (for carefully chosen values of thresholds).

\begin{table}[ht]
\centering
  \caption[NILM Metrics]{}
  \small {Home-NILM Results}
  \begin{center}
    \begin{tabular}{c|c|c}
      \hline Appliance & Threshold & Accuracy \\ \hline \hline
      Dish Washer & 0.05 & 94.6\%\\ \hline
      Microwave & 0.055 & 94.41\%\\ \hline
      Refrigerator & 0.4 & 86.58\%\\ \hline
      Washer Dryer & 0.025 & 89.97\%\\ \hline
    \end{tabular}
  \end{center}
  \label{tab:nilmmetrics}
\end{table}

\begin{figure}
    \centering
    \subfigure[]{\includegraphics[width=0.24\textwidth]{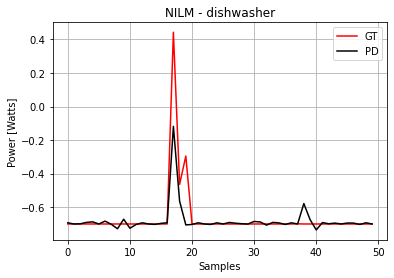}} 
    \subfigure[]{\includegraphics[width=0.24\textwidth]{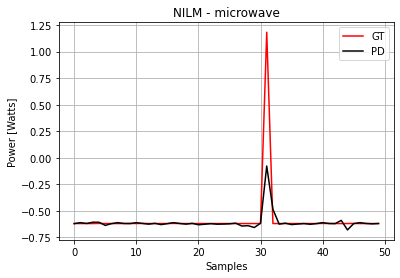}} 
    \caption{Home-NILM (a) on Dishwasher (b) on Microwave. GT is acronym for ground truth, PD is acronym for prediction by the seq2-[3]point model. }
    \label{fig:homenilm}
\end{figure}

Fig. \ref{fig:homenilm} provides a qualitative assessment of the performance of proposed DL model for home-NILM for two appliances: a dishwasher and a microwave. For each appliance, we observe that the load profile predicted by the proposed DL model closely follows the its actual load profile (note that both load profiles, i.e., actual and predicted, are in normalized form). \\

\subsubsection{Site-NILM results}
The proposed seq2-[3]point model was trained and tested on REFIT dataset in order to do site-NILM. Recall from Section III-C that the site under consideration is a Computer site with a PC and a monitor. For the classification problem (with four classes) described in Section III-C, the proposed DL was able to achieve an overall test accuracy of 81\% (where form 6,000,000 samples, 4,860,000 were correctly classified).  


        

        


\subsection{Appliance Identification Results}

Recall from Section V-B that the Wavelet appliance signatures were classified using ResNet18 and AlexNet DL models with custom FC layers, 
while the fused signatures were classified using the Densenet-121 DL model with custom FC layers. 

 

Table {\ref{tab:waveletmetrics}} compares the classification accuracy and F1 score achieved by each of the three DL models. From Table {\ref{tab:waveletmetrics}}, we learn the following: 1) the accuracy and F1 score of DenseNet model improves with a lower value of learning rate; 2) ResNet model outperformed the other two models with a classification accuracy of 88.9\% and F1 score of 88\%.

\begin{table}[ht]
\centering
  \caption[Wavelet Classification Metrics]{}
  \small {Appliance Identification Results}
  \begin{center}
    \begin{tabular}{c|c|c|c|c}
      \hline Model & Feature & Learn. Rate & Accuracy & F1 Score \\ \hline \hline
      ResNet18 & wavelets & 0.001 & 88.9\% & 88\%\\ \hline
      AlexNet & wavelets & 0.001 & 80.67\% & 81\%\\ \hline
      DenseNet-121 & fused & 0.001 & 86.17\% & 86\%\\ \hline
      DenseNet-121 & fused & 0.01 & 85.52\% & 85\%\\ \hline
    \end{tabular}
  \end{center}
  \label{tab:waveletmetrics}
\end{table}

\section{Conclusion}

This work proposed a seq2-[3]-point CNN model to solve the (home) NILM problem and site-NILM problem. We built upon the state-of-the-art (pre-trained) 2D-CNN models, i.e., AlexNet, ResNet-18, and DenseNet-121, which were trained upon two custom datasets that consist of Wavelets and STFT-based 2D electrical signatures of the appliances, to do appliance classification. We also did some basic qualitative inference about an individual appliance's health by comparing the power consumption of the same appliance across multiple homes. We achieved a maximum accuracy of 94.6\% for home-NILM, 81\% for site-NILM, and 88.9\% for appliance identification (with ResNet-based model). The low-complexity of proposed DL models makes them well-suited for edge computing for the consumer electronic devices of a smart home.

\end{document}